\documentclass[aip, jcp, twocolumn, frontmatterverbose, amssymb, amsmath, reprint]{revtex4-1}

\usepackage{graphicx}
\usepackage{hyperref}
\usepackage{multirow}
\usepackage{verbatim}

\usepackage{fontenc}

\begin{document}

\title{Unraveling the luminescence signatures of chemical defects in polyethylene}
\author{Lihua Chen}

\author{Huan Doan Tran}

\author{Chenchen Wang}

\author{Rampi Ramprasad}
\email{rampi@ims.uconn.edu}
\affiliation{Department of Materials Science and Engineering, Institute of Materials Science, University of Connecticut, 97 North Eagleville Road, Storrs, Connecticut 06269, USA}

\begin{abstract}
Chemical defects in polyethylene (PE) can deleteriously downgrade its electrical properties and performance. Although these defects usually leave spectroscopic signatures in terms of characteristic luminescence peaks, it is nontrivial to make unambiguous assignments of the peaks to specific defect types. In this work, we go beyond traditional density functional theory calculations to determine defect-derived emission and absorption energies in PE. In particular, we characterize PE defect levels in terms of thermodynamic and adiabatic charge transition levels that involve total energy calculations of neutral and charged defects. Calculations are performed at several levels of theory including those involving (semi)local and hybrid electron exchange-correlation functionals, and many-body perturbation theory. With these critical elements, the calculated defect transition levels are in excellent correspondence to observed luminescence spectra of PE, thus clarifying and confirming the origins of the observed peaks. Based on this work, a prescription with a reasonable computational expense is proposed to accurately predict and assign spectroscopic signatures of defects in other organic polymers as well.

\end{abstract}

\maketitle
\section{Introduction}
Polyethylene (PE) is an important insulation material that has found widespread use in electrical applications including transmission line cables and capacitors. \cite{peacock2000handbook} The electrical performance of PE over the long term is affected by impurities and chemical defects that are originally part of the material, as well as those that are created progressively with time. Such defects can introduce charge carrier (or defect) states within the band gap of PE, can act as “traps” and sources of charge carriers, catalyze further damage, and can deleteriously affect the overall conduction behavior of the insulator.\cite{patsch1990space,ieda1997space,dissado1997role,ceresoli2004trapping,mazzanti2005electrical} It is thus critical that a firm understanding of the nature of such defects be obtained.

\begin{figure}[t]
\centering
\includegraphics[width=8cm] {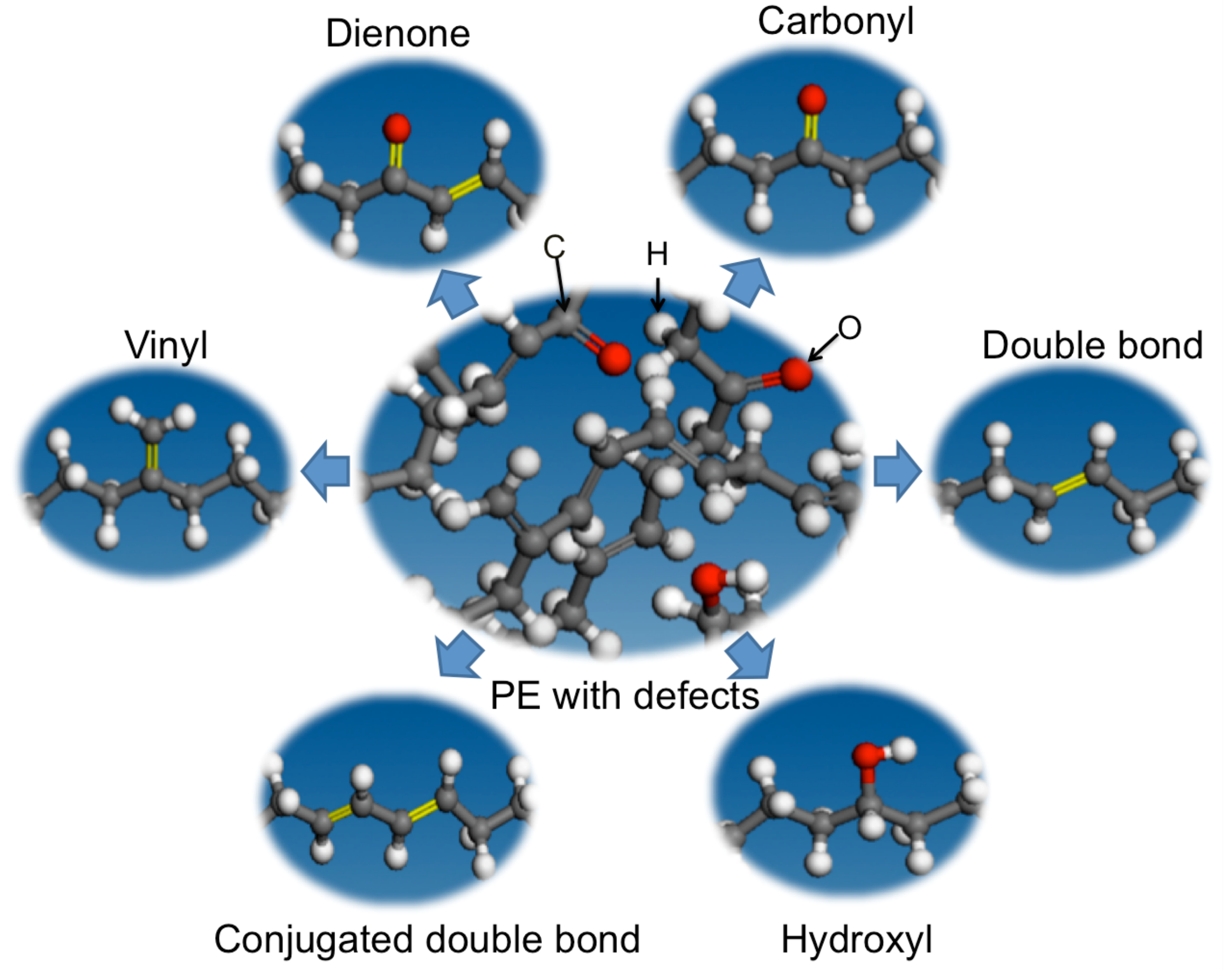}
\caption{Six typical defects that occur in polyethylene are considered in this work, e.g., carbonyl, dienone, hydroxyl, double bond, conjugated double bond and vinyl. The red, white and grey spheres are O, H and C atoms, respectively. }\label{fig:defect}
\end{figure}

The best evidence for the presence of chemical defects (especially in PE), and a knowledge of their type, is provided by a variety of luminescence measurements, including electroluminescence, photoluminescence and chemiluminescence.\cite{taleb2013modeling, allen1977identification, teyssedre2002characterisation} Defects, depending on their type and the specific details of the energetic placement of their defect levels, lead to characteristic emission signatures. Such measurements have led to the identification of a plethora of defects in PE, the chief among them being the carbonyl, dienone, vinyl, hydroxyl, etc., as illustrated in Fig. \ref{fig:defect}. Nevertheless, assignments of the luminescence emission bands are never straightforward, and an alternate check of whether a particular defect will lead to a particular emission peak or band is highly desired.

Such an alternate route to identify emission signatures can be provided by first principles computations, e.g., those based on density functional theory (DFT).\cite{DFT1, DFT2} Indeed, DFT and beyond-DFT computations have played a critical role in successfully unraveling the repercussions and signatures of defects in inorganic semiconductors and insulators in the past.\cite{RevModPhys.86.253, PhysRevLett.108.066404, yan2014origins} In the case of PE, such work is in a state of infancy.\cite{bealing2013atomistic, huzayyin2010density, huzayyin2010quantum, meunier2001molecular, anta2002models} Past DFT work on PE is dominated by the use of (semi)local exchange-correlation functionals to treat the quantum mechanical part of the electron-electron interaction within a single-particle framework. Defect states are identified as the one-electron Kohn-Sham eigenenergies.\cite{huzayyin2010density, huzayyin2010quantum, meunier2001molecular, anta2002models} While this approach does provide a qualitative picture of the defect-derived energy levels, there are two fundamental drawbacks. The first one is that the one-electron levels of DFT have no physical meaning as they relate to the energy levels of a fictitious set of non-interacting electrons.\cite{DFT2} The second issue relates to the band gap problem of traditional DFT.\cite{seidl1996generalized} It is well-known that the band gap of insulators are significantly underestimated, which place uncertainties on the energetic location of the defect energy levels within the band gap as well as the position of the conduction and valence band edges.\cite{hybertsen1985first,gruning2006density} These factors clearly call into question the emission energies (i.e., the energy difference between defect levels and the appropriate band edges) derived from such calculations.

In the present work, we go beyond the single-particle picture of defect levels in PE. Defect levels are characterized in terms of thermodynamic and adiabatic charge transition levels that involve total energy differences of charged and neutral defect calculations. It is well-known that total energy differences provide a formally correct and quantitatively better description of energy level differences in molecules and solids (including defect level placements). \cite{PhysRevLett.108.066404, yan2012role, yan2014origins} Exchange-correlation functionals more advanced than the (semi)local ones used in the past are considered in the present work to critically assess the role of the level of theory adopted. Furthermore, many-body perturbation theory computations are undertaken to accurately determine the band gap of PE and its conduction and valence band edge positions. 

The combination of computations undertaken here, including the determination and usage of charge transition levels (rather than the one-electron levels) and the accurate band edge and gap descriptions allow us to directly and quantitatively connect with available luminescence data for PE. The present work thus leads to a clear and compelling picture of defect states in PE, and provides a systematic and reliable pathway for determining the luminescence signatures of various defects in PE as well as other insulators.

\section{Models and methods}\label{sec:model_method}

\subsection{Models}

Perfect crystalline PE consists of chains of singly-bonded $-$CH$_2-$ groups, of which all the carbon atoms display $sp^3$ hybridization. In this work, $1\times 2\times3$ supercells of crystalline PE, containing of 24 $-$CH$_2-$ groups, are constructed with and without defects as the models for our calculations.
Six kinds of defects, illustrated in Fig. \ref{fig:defect}, are typically considered to exist in PE chains. A \textit{carbonyl} defect (C=O) is created when the two H atoms of a $-$CH$_2-$ group are replaced by an O atom doubly bonded to the central C atom, which now displays $sp^2$ hybridization. If the two H atoms of --CH$_{2}$-- are replaced by a $=$CH$_2$ group, we have a \textit{vinyl} defect. When each of two adjacent C atoms loses one H atom, a \textit{double bond} (--CH=CH--) may be formed between them. The combination of a double bond and a carbonyl is \textit{dienone} (--CH=CH--C=O). In the case of  a \textit{hydroxyl} defect, a H atom is replaced by an $-$OH group. Along the PE chains, an alternating pattern of single and double bonds leads to  \textit{conjugated double bonds} (--CH=CH--CH=CH--). 

\subsection{DFT computations}
Our DFT calculations were performed with the Vienna {\it ab-initio} simulation package (VASP).\cite{vasp2,vasp3,vasp4} The (folded) Brillouin zone of the supercell is sampled by a $2\times 2\times2$ {\bf k}-point Monkhorst-Pack mesh \cite{monkhorst} while plane waves with kinetic energy up to 400 eV were included in the basis set. van der Waals interactions are considered within the scheme developed by Tkatchenko and Scheffler (TS).\cite{tkatchenko2009accurate} 
Relaxed geometries were obtained  with Perdew-Burke-Ernzerhof (PBE) exchange-correlation (XC) functional, \cite{perdew1996generalized} which were then used to compute the thermodynamic and optical transition levels with the PBE and the Heyd-Scuseria-Ernzerhof (HSE)\cite{HSE06:1, HSE06:2} XC functionals, and  the self-consistent many-body perturbation theory GW method. 
In the case of the HSE XC functional, we used the HSE06 version with the mixing coefficient $\alpha = 0.25$ and the screening parameter $ \omega = 0.2$ \AA$^{-1}$. The band gap ($E_{\rm g}$) of perfect PE calculated with PBE, HSE06 and GW methods are 7.00 eV, 8.39 eV, and 8.95 eV, respectively, with the GW result being closest to the experimental value (8.8 eV). \cite{less1973intrinsic} 

\subsection{Defect levels theory}
Defect levels are probed by enforcing electronic transitions, such as those involving a transfer of an electron from (to) a defect state to (from) the conduction band minimum (the valence band maximum) of PE. Such transitions can occur in two ways, depending on the time scale of the measurement process. Thermodynamic transitions (involving long time scales) will involve initial and final charge states at the respective equilibrium geometries.
Adiabatic (or optical) transitions will occur at much shorter time scales, and will involve the final charge state at the equilibrium geometry of the initial charge state.

\begin{figure}[t]
\begin{center}
\includegraphics[width=8.5cm] {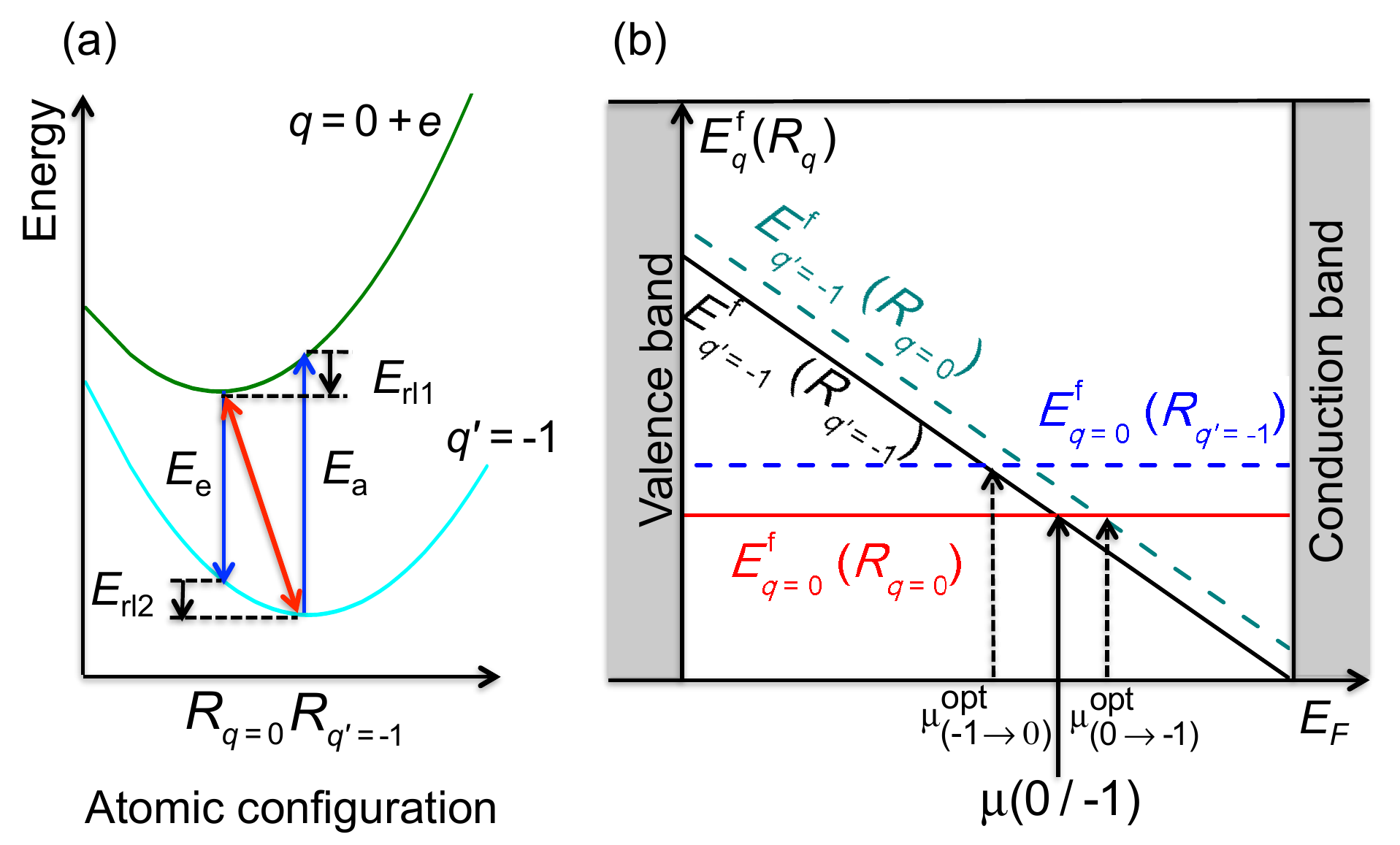}
\caption{(Color online) 
(a) A configuration coordinate diagram for a defect at charge states $-1$ and $0$ which correspond to two equilibrium configurations $R_{q'=-1}$ and $R_{q=0}$, shown as the minima of the potential energy surfaces (cyan and olive lines). The {\it instant} transition from $0$ state to $-1$ state gains the emission energy $E_e$ while the configuration $R_{q = 0}$ remains unchanged. After a certain amount of time, the $-1$ state evolves to $R_{q'=-1}$, gaining $E_{\rm rl2}$ from the relaxation process. Similarly, the transition from the $-1$ state to the $0$ state by absorbing $E_a$ happens instantly before the system can relax, gaining $E_{\rm rl1}$. (b) Formation energy $E^{\rm f}_{q}(R_{q}$) as a function of Fermi energy ($E_{F}$) for a defect that has two stable charge states: 0 and $-1$. $E^{\rm f}_{q}$($R_{q'}$) refers to the formation energy of the defect in charge state $q$ with the equilibrium configuration of charge state $q'$ ($R_{q'}$), where $q$ and $q'$ are 0 or $-1$. 
Solid lines are the formation energies of the relaxed defects in each charge state. While the formation energies for the defect with frozen atomic configuration of initial charge state are presented with dashed lines. Both thermodynamic transition level $\mu$(0/$-$1) and the relevant optical transition levels ($\mu^{\rm opt}_{(0\to -1)}$ and $\mu^{\rm opt}_{(-1\to 0)}$) are shown. }\label{fig:opticallevel}
\end{center}
\end{figure}

A formally correct approach to determine defect levels is via total energy differences of PE with defects at initial and final charge states.\cite{van2004first,RevModPhys.86.253, PhysRevLett.108.066404} The local equilibrium atomic configurations are different for the defects in different charge states, \cite{RevModPhys.86.253} as illustrated in Fig. \ref{fig:opticallevel} (a). A configuration--coordinate diagram of a defect in two charge states: 0 ($q$) and $-1$ ($q'$) is shown in Fig. \ref{fig:opticallevel} (a), in which the minima of the potential energy surfaces represent the ground state of the defect in charge state $q=0$ and $q'=-1$. In the case of thermodynamic transition, the defect has enough time to relax from its initial ground state at $q=0$ into its new ground states at $q'=-1$ or vice versa, as shown using red double arrowhead line in Fig. \ref{fig:opticallevel} (a). In order to get this charge transition level, the formation energies $E^{\rm f}_{q}$($R_{q}$) as a function of Fermi energy ($E_{F}$) for this defect at charge states $q=0$ and $q'=-1$ are computed and shown in Fig. \ref{fig:opticallevel} (b), where $R_{q}$ represents the equilibrium structures in charge state $q$. Based on  Fig. \ref{fig:opticallevel} (a), the thermodynamic transition level $\mu$(0/$-$1) corresponds to the crossover point between  $E^{\rm f}_{q=0}$($R_{q=0}$) and $E^{\rm f}_{q=-1}$($R_{q'=-1}$) (solid lines in Fig. \ref{fig:opticallevel} (b)). In general, the thermodynamic transition levels $\mu (q/q')$ is the Fermi energy at which defects in two different charge states $q$ and $q'$ are at thermodynamic equilibrium and is given by \cite{van2004first,RevModPhys.86.253, PhysRevLett.108.066404} 
\begin{equation}
\mu (q/q')=\frac{E^{\rm f}_{q}(R_q)-E^{\rm f}_{q'}(R_{q'})}{q-q'}. 
\label{eq:thermal}
\end{equation}
Here, $E^{\rm f}_{q}(R_q)$ is the formation energy of the $q$-charged defect at its equilibrium structure $R_{q}$, which can be obtained from DFT calculations. The Fermi energy is taken from valence band maximum (VBM) to conduction band minimum (CBM) of defect-free PE. In this work, all energies are referenced to the averaged electrostatic potential of PE far away from the defect location. Due to the unavailability of the total energy with GW method, the GW defect levels are corrected by including a term $\Delta E_{\rm VBM}= {\rm VBM}_{\rm GW}- {\rm VBM}_{\rm HSE06}$ for the VBM shift of defect-free PE, expressed as 
\begin{equation}\label {gw-th}
(\mu (q/q'))_{\rm GW}=(\mu (q/q'))_{\rm HSE06}+ \Delta E_{\rm VBM}.
\end{equation}
The subscripts (HSE06 and GW) are refer to the method used to calculate the relevant quantities. 

\begin{figure}[t]
\begin{center}
\includegraphics[width=8.5cm] {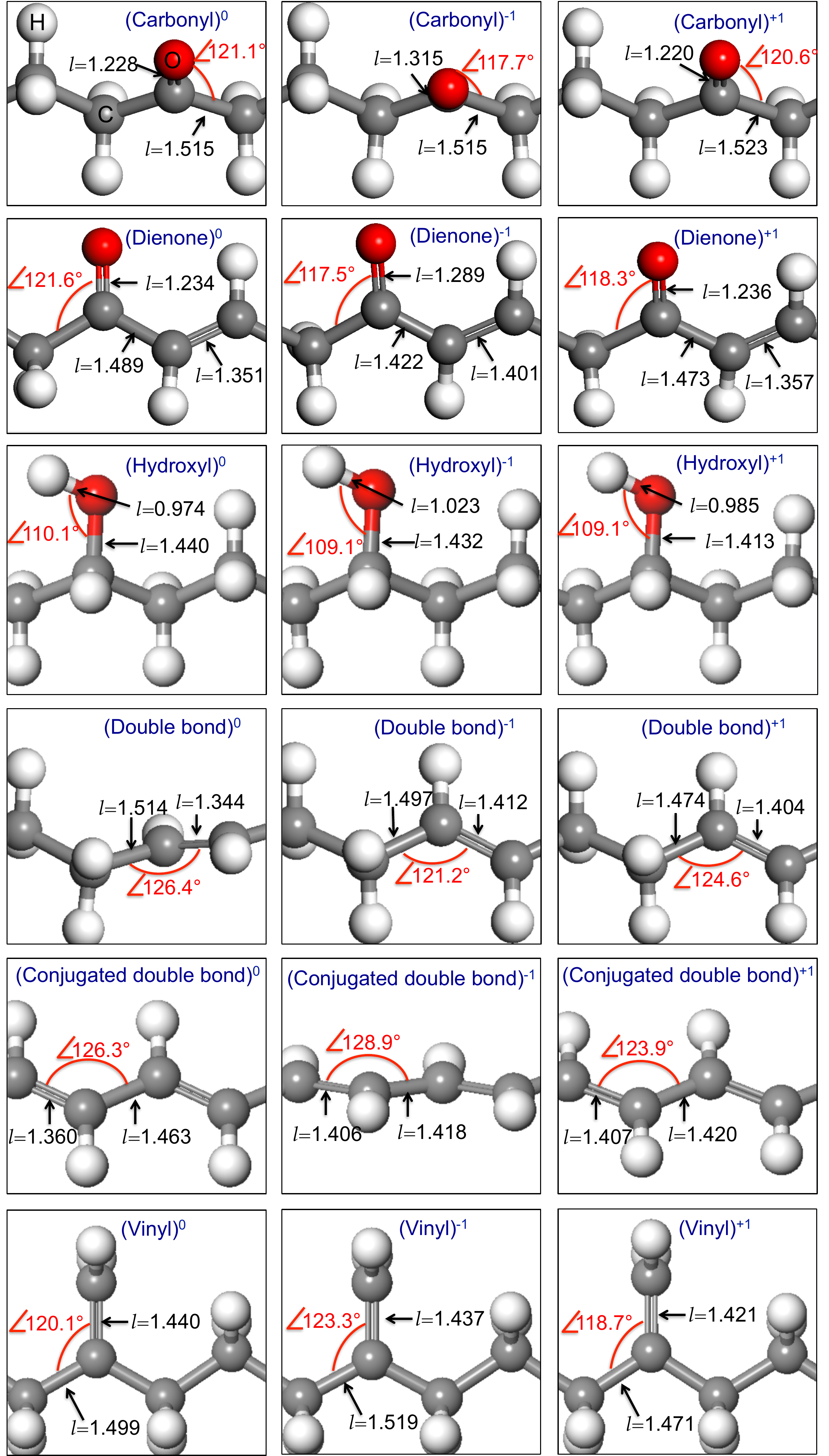}
\caption{(Color online) Side view of the relaxed PE model with defects at different charge states [denoted by (defect)$^{q}$, $q$ = 0, $-1$ and +1], optimized with PBE functional. The red, white and grey spheres are, respectively, O, H and C atoms. The bond angles ($\angle$) and bond lengths ($l$, given in ${\rm \AA} $) are shown in red and black, respectively.}\label{fig:geo}
\end{center}
\end{figure}

In the case of optical transition, the atomic configuration of the defect at the initial charge state $q$ is fixed even though charge transition exists, i.e., it is a vertical transition which is shown using a blue arrowhead line in Fig. \ref{fig:opticallevel} (a). Because the optical transition depends on the direction of charge transfer, two kinds of optical energies are possible, as shown in Fig. \ref{fig:opticallevel} (a): emission energy ($E_{\rm e}$) for charge transfer from initial state $0$ to final state $-1$  and absorption energy ($E_{\rm a}$) for the reverse process. The method to determine the optical transition level  is similar to the previous case, but the energy of the final state is computed using the equilibrium structure of the initial state, such as $\mu^{\rm opt}_{(0\to -1)}$ and $\mu^{\rm opt}_{(-1\to 0)}$ in Fig. \ref{fig:opticallevel} (b).  Therefore, the optical transition level from initial state $q$ to final state $q'$ is defined as 
 \begin{equation}
\mu^{\rm opt}_{( q \rightarrow q')}=\frac{E^{\rm f}_{q'} (R_q)- E^{\rm f}_{q}(R_q)}{(q'-q)},
\label{eq:opt}
\end{equation}
where $E^{\rm f}_{q'}(R_q)$ is the defect formation energy in the charge state $q'$ corresponding to the equilibrium structure $R_{q}$ of the initial charge state $q$. 
The corresponding GW optical transition level is then expressed as 
\begin{equation}
(\mu^{\rm opt}_{( q \rightarrow q')})_{\rm GW}=(\mu^{\rm opt}_{( q \rightarrow q')})_{\rm HSE06}+\Delta E_{\rm VBM}.\label {gw-op}
\end{equation}

\section{Results and Discussions}
\subsection{Defect geometries}

 The relaxed structures of PE with defects at different charge states $q$ [denoted by (defect)$^{q}$, $q$ = 0, $-1$ and +1] are shown in Fig. \ref{fig:geo}. The variation of geometry parameters (bond angle $\angle$ and bond length $l$) of (defect)$^{q}$ indeed shows that the equilibrium structures corresponding to different charge states are different. For example, the C=O bond length of carbonyl increases from 1.228 ${\rm \AA} $ for the neutral state to 1.315 ${\rm \AA} $ for the $-1$ charge state, because the extra electron is localized at the C=O bond and leads to bond weakening. Whereas, compared to the  (carbonyl)$^{0}$, the decrease of the C=O bond length in (carbonyl)$^{+1}$ is due to the loss of an electron.  As another example, we note that  a rotation of about 90 degrees can be observed for C=C part of (double bond)$^{-1}$ and (double bond)$^{+1}$ compared to the neutral case. This is because both positive and negative charges can weaken the C=C bond and make the PE chain free to rotate. All these relaxations for (defect)$^{q}$ directly cause the broad luminescence peaks in experiments.  Therefore, in this work, all levels ranging from $\mu^{\rm opt}_{( q \to q')}$, which is relevant to the emission energy, to $\mu( q/ q')$ are used to compare with experimental results.

\begin{figure*}[t]
\includegraphics[width=16cm]{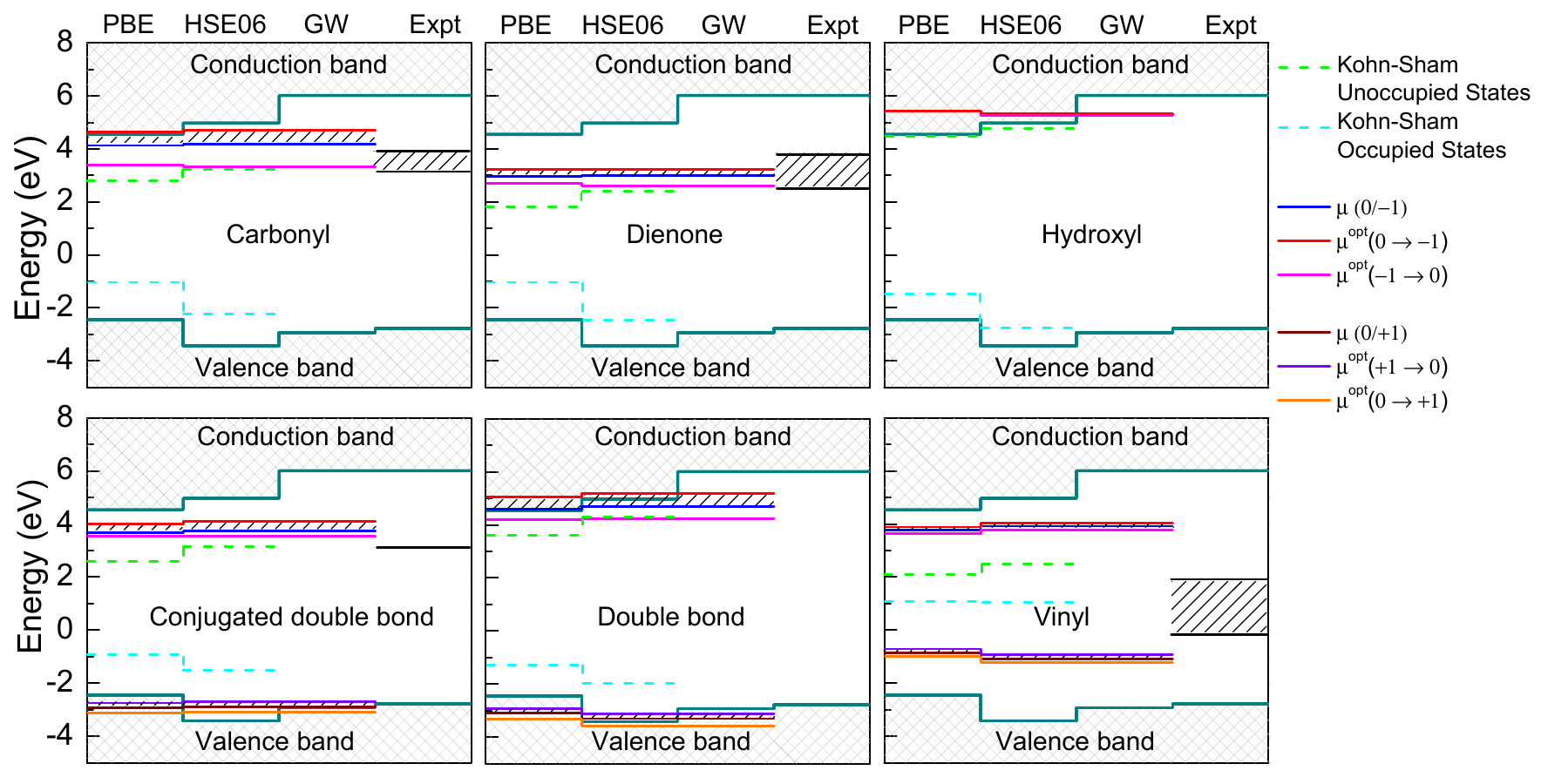}
\caption{(Color online) The thermodynamic (i.e., $\mu$(0/$-$1) and $\mu$(0/+1)) and optical (i.e., $ \mu^{\rm opt}_{(0 \to -1)}$, $ \mu^{\rm opt}_{(-1 \to 0)}$, $ \mu^{\rm opt}_{(0 \to +1)}$, and $ \mu^{\rm opt}_{(+1 \to 0)}$) transition levels are represented with blue, wine, red, magenta, violet and orange solid lines, respectively. The VBM and CBM are given with respect to the average electrostatic potential.  The possible transition levels which lead to optical emissions are represented by the shaded region. The experimental levels are determined by the difference between the CBM and experimental emission energies from Ref. \onlinecite{laurent2013charge,allen1977identification,ito2002origins}. For convenience,  the experimental CBM of defect-free PE is aligned with the GW CBM.  For completeness, the Kohn-Sham unoccupied and occupied states are also shown using green and cyan dashed lines, respectively.} \label{fig:trans}
\end{figure*}

\subsection{Charge transition levels}
In order to characterize the defect levels in PE, which are responsible for the observed luminescence signatures, the thermodynamic and optical transition levels for six typical defects in PE have been calculated. The results obtained from various methods (solid lines) and available experimental data are shown in Fig. \ref{fig:trans}. As can be seen, both (0/$-$1) and (0/+1) charge transition levels are obtained for double bond, conjugated double bond and vinyl defects. For the carbonyl, dienone and hydroxyl defects, just the (0/$-$1) charge transition level exists, indicating that these three defects at positive charge state are not stable within the band gap of PE. The associated thermodynamic (i.e., $\mu$(0/$-$1) and $\mu$(0/+1)) and optical (i.e., $ \mu^{\rm opt}_{(0 \to -1)}$, $ \mu^{\rm opt}_{(-1 \to 0)}$, $ \mu^{\rm opt}_{(0 \to +1)}$, and $ \mu^{\rm opt}_{(+1 \to 0)}$) transition levels are represented with blue, wine, red, magenta, violet and orange solid lines, respectively.

\begin{table*}[ht]
\begin{center}
\caption{ The computed emission energies (ranging from ${\rm CBM} - \mu^{\rm opt}_{(q \rightarrow q')}$ to ${\rm CBM}- \mu(q/q')$) and available experimental results for defects involved in different charge transfer processes, given in eV. PL, EL, and OL stand for photoluminescence, electroluminescence, and other luminescence, respectively.} \label{table:RE}
\begin{tabular}{p{2.5cm}c c c c c c c c c c c c c c }
\hline
\hline
Defects &\multicolumn{3}{c}{Charge transfer } & \multicolumn{3}{c}{Optical emission energies}&\multicolumn{3}{c}{Experimental values}\\

& $q$ & $\rightarrow$ &$q'$ & PBE & HSE06 & GW & PL & EL & OL\\
\hline
Carbonyl&0&$\rightarrow$& $-1$&$-$ & $-$ & $1.32$ -- 1.84&$2.06$ -- $3.10$ (Ref. \onlinecite{allen1977identification})& $2.92$ (Ref. \onlinecite{laurent2013charge})&$-$\\

Dienone & 0& $\rightarrow$& $-1$ & $1.31$ -- 1.57& $1.75$ -- 1.99& $2.36$ -- 3.04& $2.25$ -- $3.54$ (Ref. \onlinecite{allen1977identification}) &$2.43$ (Ref. \onlinecite{laurent2013charge})&$-$ \\
Hydroxyl&0&$\rightarrow$& $-1$ & $-$ & $-$ & $0.70$ & $-$ & $-$ & $-$\\
Conjugated double bond&0&$\rightarrow$& $-1$ & $0.53$ -- 0.85& $0.91$ -- 1.23& $1.96$ -- 2.28& $2.90$ (Ref. \onlinecite{ito2002origins}) & $-$ & $-$\\

Double bond&0&$\rightarrow$& $-1$& $-$ & $-$ & 0.92 -- 1.42 & $-$ & $-$ &$-$ \\

Vinyl&+1&$\rightarrow$& $0$ & 5.26 -- 5.40& 5.91 -- 6.05 &6.96 -- 7.10 & $-$& $-$& \multirow{1}{*}{$4.13$ -- $6.20$ (Ref. \onlinecite{allen1977identification})}\\
\hline
\hline
\end{tabular}
\end{center}
\end{table*}

Based on  the configuration-coordinate diagram in Fig. \ref{fig:opticallevel} (a), we find that the $ \mu^{\rm opt}_{(0 \to -1)}$ and $ \mu^{\rm opt}_{(+1 \to 0)}$ are directly relevant to luminescence emissions. However, because of the broad peak observed in experiments, the levels from $ \mu^{\rm opt}_{(0 \to -1)}$ to $\mu$(0/$-$1) or from $ \mu^{\rm opt}_{(+1 \to 0)}$ to $\mu$(0/+1) are used to compare with experimental levels, as shown by the shaded region in Fig. \ref{fig:trans}. The experimental levels described here, are determined by the difference between the CBM and experimental emission energies from Ref. \onlinecite{laurent2013charge,allen1977identification,ito2002origins}. To clarify the interpretation, the experimental CBM of defect-free PE is aligned with the GW CBM.

From Fig. \ref{fig:trans}, we observe that the GW optical transition levels $ \mu^{\rm opt}_{(0 \to -1)}$ for carbonyl, dienone, and conjugated double bond defects are in good agreement with the experimental luminescence levels, which suggests that the luminescence of such defects originate from a transition of an electron from the conduction band to a neutral defect, i.e., $({\rm defect})^{0} + e \rightarrow ({\rm defect})^{-1}$. On the other hand, the GW $\mu^{\rm opt}_{( +1 \to 0)}$  of vinyl defect is closer to the experimental result compared to $\mu^{\rm opt}_{( 0 \to -1)}$. This implies that the transition of electrons from the conduction band to the charge state $+1$ of vinyl, i.e., $ ({\rm CH=CH_{\rm 2}})^{+1} + e \to({\rm CH=CH_{\rm 2}})^{0}$ is most likely the origin of the luminescence center caused by the vinyl defect. However,  a small discrepancy still remains between computational and experimental results for the vinyl defect. The reason for this disagreement could be because there are uncertainties in the experimental data, or more likely, because the present calculations considered defects in a crystalline PE environment. The latter aspect may affect the vinyl defect the most as this is the bulkiest of defects considered here. Nevertheless, overall, we are able to achieve an acceptable correspondence with available experimental luminescence data for defects in PE.

It is also worth noting a few curious details related to the computed results. When referenced to the average electrostatic potential, the charge transition levels calculated with PBE agree well with those based on the HSE06 and the GW method, as shown in Fig. \ref{fig:trans}. This aspect, as already pointed out earlier, \cite{PhysRevLett.108.066404} reveals that the charge transition levels are captured well by the PBE functional, if a suitable reference energy is used. 

Unlike the charge transition levels, a remarkable variation (with the level of theory used) of the band edges of PE is observed in Fig. \ref{fig:trans}, which directly impacts the comparison with experimental emission energies (which are the energy differences between the defect transition levels and the appropriate band edge energies). To show the role of band edges, the emission energies of defects computed with the PBE, HSE06 and the GW method are listed in Table \ref{table:RE}, together with available experimental results. The consistency of GW values and experimental emission energies indicates that an accurate prediction of band edges is key to the correct estimation of the emission energies.

Based on our results, we propose  the following systematic procedure to compute emission and absorption energies accurately: (1) calculation of $\mu^{\rm opt}_{( q \to q')} $ and $\mu( q/ q')$ using the PBE functional; (2) calculation of VBM and CBM of pure PE with the GW method; (3) correction of $\mu^{\rm opt}_{( q \to q')} $ and $\mu( q/ q')$ with VBM values from the GW method (using Eqs. (\ref{gw-th}) and (\ref{gw-op})) by using the average electrostatic potential as the energy reference; and finally (4) the determination of the emission energies as the energy differences between appropriate  defect transition levels and band edges. Generally, this procedure can be used to predict and assign spectroscopic signatures of defects in other organic polymers as well.
\subsection{Kohn-Sham levels}
To complete the discussion, and to put our work in the context of the existing literature pertaining to defects and defect states in PE, we also include the one-electron Kohn-Sham defect levels for various defects in PE in Fig. \ref{fig:trans}. These are shown using green (unoccupied states) and cyan (occupied states)  dashed lines in Fig. \ref{fig:trans}. While the Kohn-Sham defect levels are in reasonable and qualitative agreement with the charge transition levels, some important observations should be made. First, even when the PBE and HSE06 average electrostatic potentials are aligned, the one-electron levels do not line up (unlike the charge transition levels). This indicates that the one-electron levels are strongly dependent on the level of theory used. Second, the emission energies that one would arrive at purely using the one-electron energies and the band edges of the same theoretical treatment would be in stark disagreement with experiments. Thus, interpreting defect spectroscopic signatures using Kohn-Sham levels should be performed with caution.

\section{Summary}
In summary, we have presented a detailed first-principles computational study of chemical defects in polyethylene (PE) and have quantitatively determined defect-derived emission and absorption energies. A recipe for accurately computing and assigning luminescence peaks for materials containing defects is proposed. This procedure involves going beyond traditional density functional theory calculations within the one-electron framework. It is concluded that defect levels should be determined in terms of thermodynamic and adiabatic charge transition levels (rather than the one-electron Kohn-Sham levels) that involve total energy calculations of neutral and charged defects, and that band edges should be determined accurately using many-body perturbation theory. As a consequence of these steps, this work offers a quantitative and satisfactory agreement between computed and measured emission/absorption energies for polyethylene containing chemical defects. It is suggested that this computational scheme may be used to interpret the luminescence data of other organic polymers as well.

\section{Acknowledgement}
This work is supported by a Multidisciplinary University Research Initiative (MURI) grant from the Office of Naval Research.

\bibliographystyle{apsrev4-1}
\bibliography{references}

\end{document}